\documentclass[aps,pre,reprint,groupedaddress]{revtex4-2}

\usepackage{graphicx}
\graphicspath{{./figs/}} 
\usepackage{bm}

\usepackage{cmap} 
\usepackage[utf8]{inputenc}
\usepackage[T1]{fontenc}
\usepackage[english]{babel}
\usepackage{mathptmx}
\usepackage{amsmath,amssymb,mathtools}
\mathtoolsset{%
above-intertext-sep = - 2pt,
below-intertext-sep = - 4pt,
}

\usepackage[colorlinks=true,linkcolor=blue,citecolor=blue,urlcolor=blue,unicode]{hyperref}

\newcommand*{\txtsubscript}[2]{{#1}_{\text{#2}}}  

\newcommand*{\const}{\mathrm{const}}  
\newcommand*{\ppartial}{\partial\!}
\let\dif\undefined
\newcommand*{\dif}{\mathop{}\!\mathrm{d}}      

\newcommand*{\bigO}{\mathrm{O}} 

\newcommand*{\Vtot}{\txtsubscript{V}{t}}
\newcommand*{\Atot}{\txtsubscript{A}{t}}

\newcommand*{\ab}{\alpha\beta}
\newcommand*{\bg}{\beta\gamma}
\newcommand*{\ag}{\alpha\gamma}

\newcommand*{\Omegaa}{\txtsubscript{\Omega}{a}}
\newcommand*{\Omegac}{\txtsubscript{\Omega}{c}}
\newcommand*{\Omegaac}{\txtsubscript{\Omega}{a,c}}

\newcommand*{\omegaac}{\txtsubscript{\omega}{a,c}}

\newcommand*{\Va}{V^\alpha}
\newcommand*{\Vb}{V^\beta}

\newcommand*{\Vaa}{\Va_\text{\!a}}
\newcommand*{\Vac}{\Va_\text{\!c}}
\newcommand*{\Vaac}{\Va_\text{\!a,c}}
\newcommand*{\Vba}{\Vb_\text{\!a}}
\newcommand*{\Vbc}{\Vb_\text{\!c}}
\newcommand*{\Vbac}{\Vb_\text{\!a,c}}   

\newcommand*{\Aab}{A^{\ab}}
\newcommand*{\Abg}{A^{\bg}}
\newcommand*{\Aag}{A^{\ag}}

\newcommand*{\Aaba}{\Aab_\text{a}}
\newcommand*{\Aaga}{\Aag_\text{a}}
\newcommand*{\Abga}{\Abg_\text{a}}
\newcommand*{\Aabc}{\Aab_\text{c}}
\newcommand*{\Aagc}{\Aag_\text{c}}
\newcommand*{\Abgc}{\Abg_\text{c}}
\newcommand*{\Aagac}{\Aag_\text{a,c}}
\newcommand*{\Abgac}{\Abg_\text{a,c}}

\newcommand*{\sgmab}{\sigma^{\ab}}
\newcommand*{\sgmaba}{\sgmab_\text{a}}
\newcommand*{\sgmabc}{\sgmab_\text{c}}
\newcommand*{\sgmabac}{\sgmab_\text{a,c}}
\newcommand*{\sgmag}{\sigma^{\ag}}
\newcommand*{\sgmbg}{\sigma^{\bg}}
\newcommand*{\sgmg}{\sigma^{\gamma}}
\newcommand*{\Dsgmg}{\Delta\sgmg}
\newcommand*{\Gab}{\Gamma^{\ab}}
\newcommand*{\Gaba}{\Gab_\text{\!a}}
\newcommand*{\Gabc}{\Gab_\text{\!c}}
\newcommand*{\Gabac}{\Gab_\text{\!a,c}}
\newcommand*{\Gbg}{\Gamma^{\bg}}
\newcommand*{\Gag}{\Gamma^{\ag}}     

\newcommand*{\na}{n^{\alpha}}
\newcommand*{\nb}{n^{\beta}}
\newcommand*{\chia}{\chi^\alpha}
\newcommand*{\nabndl}{\na_{\infty}}

\newcommand*{\chiabndl}{\chia_{\infty}}
\newcommand*{\pabndl}{p^\alpha_{\infty}}    

\newcommand*{\sgmabbndl}{\sigma^{\alpha\beta}_{\infty}}
\newcommand*{\sgmbgbndl}{\sigma^{\beta\gamma}_{\infty}}
\newcommand*{\sgmagbndl}{\sigma^{\alpha\gamma}_{\infty}}
\newcommand*{\Dsgmgbndl}{\Delta\sigma^{\gamma}_{\infty}}
\newcommand*{\Gabbndl}{\Gamma^{\alpha\beta}_{\!\infty}}
\newcommand*{\Gbgbndl}{\Gamma^{\beta\gamma}_{\!\infty}}
\newcommand*{\Gagbndl}{\Gamma^{\alpha\gamma}_{\!\infty}}

\newcommand*{\fbndl}{f_{\!\infty}}       

\newcommand*{\ra}{\txtsubscript{r}{\!a}}
\newcommand*{\rc}{\txtsubscript{r}{\!c}}
\newcommand*{\rac}{\txtsubscript{r}{\!a,c}}

\newcommand*{\Ra}{\txtsubscript{R}{a}}
\newcommand*{\Rc}{\txtsubscript{R}{c}}
\newcommand*{\Rac}{\txtsubscript{R}{a,c}}

\newcommand*{\La}{\txtsubscript{L}{a}}
\newcommand*{\Lc}{\txtsubscript{L}{c}}
\newcommand*{\Lac}{\txtsubscript{L}{a,c}}

\newcommand*{\thetaa}{\txtsubscript{\theta}{a}}
\newcommand*{\thetac}{\txtsubscript{\theta}{c}}
\newcommand*{\thetaac}{\txtsubscript{\theta}{a,c}}
\newcommand*{\thetabndl}{\theta_\infty}
\newcommand*{\thetas}{\txtsubscript{\theta}{s}}

\newcommand*{\kappaa}{\txtsubscript{\kappa}{a}}
\newcommand*{\kappac}{\txtsubscript{\kappa}{c}}
\newcommand*{\kappaac}{\txtsubscript{\kappa}{a,c}}

\newcommand*{\kappaappa}{\txtsubscript{\varkappa}{a}}
\newcommand*{\kappaappc}{\txtsubscript{\varkappa}{c}}

\newcommand*{\kappabndl}{\kappa_{\infty}}

\newcommand*{\mubndl}{\mu_\infty}
                                              
\newcommand*{\TolmanLength}{\txtsubscript{\delta}{T}}

\newcommand*{\ma}{\txtsubscript{m}{a}}
\newcommand*{\mc}{\txtsubscript{m}{c}}
\newcommand*{\mac}{\txtsubscript{m}{a,c}}

\newcommand*{\Lambdabndl}{\Lambda_\infty}

\begin{document}

\title[Line tension from dual-geometry sessile droplet measurements]{Line tension from dual-geometry sessile droplet measurements: Combining contact-angle size-dependence data for axisymmetric and cylindrical droplets to determine the line tension}

\author{Dmitry V. Tatyanenko}
\email{d.tatyanenko@spbu.ru}

\author{Konstantin D. Apitsin}

\affiliation{Department of Statistical Physics, Faculty of Physics, Saint Petersburg State University\\ 7--9 Universitetskaya nab., St.~Petersburg, 199034, Russia}

\date{\today}

\begin{abstract}
We perform a thermodynamic analysis of various contributions to the size dependence of the contact-angle cosine for both axisymmetric and cylindrical sessile droplets. This shows that a commonly used method to determine the line tension from the slope of the contact-angle cosine dependence on the three-phase contact-line curvature (for axisymmetric droplets) provides a certain combination of the line tension, the adsorptions at the three interfaces, and the macroscopic contact angle. To extract the contribution related to the contact line in the leading order and determine the line tension, we propose a simple technique using droplet-size dependences of the contact angle for axisymmetric and cylindrical droplets under the same conditions.

\bigskip

\noindent
\href{https://doi.org/10.1103/PhysRevE.111.035503}{Phys. Rev.~E \textbf{111}, 035503 (2025)}. DOI: \href{https://doi.org/10.1103/PhysRevE.111.035503}{10.1103/PhysRevE.111.035503}
\qquad 
\copyright 2025 American Physical Society
\end{abstract}


\maketitle

\section{\label{sec:introduction}Introduction}

\begin{figure}[b]
\includegraphics[width=8.2cm]{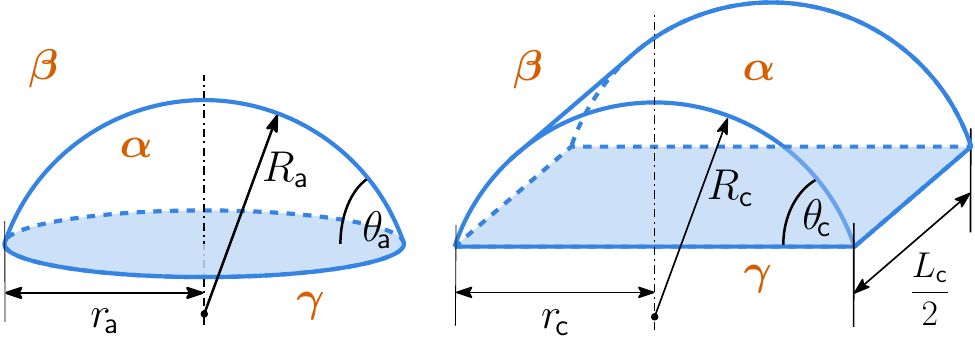}

\vspace{0.1cm}
\hspace{0.4 cm} (a) \hspace{3.3 cm} (b) \hspace{0.7 cm}

\caption{\label{fig:droplets} Axisymmetric (a) and cylindrical (b) sessile droplets. $R$ is the curvature radius of the liquid--gas interface, $\theta$ is the contact angle, $r$ is the radius (half-width) of the droplet on the substrate. Subscript ``a'' marks quantities for axisymmetric, ``c''~--- for cylindrical droplets. Phases: $\alpha$~--- liquid, $\beta$~--- gas (vapor), $\gamma$~--- solid substrate.}
\end{figure}

The contact angle of a large sessile droplet can be found from the well-known classical Young equation~\cite{Young-PhilTransRSocLond-1805}
\begin{equation}\label{cYe}
\sgmab \! \cos\thetabndl = \sgmbg - \sgmag
\end{equation}
with $\sigma$ the surface tension of the corresponding interface and $\thetabndl$ the macroscopic contact angle; the superscripts $\alpha$, $\beta$, and $\gamma$ mark the liquid, gaseous, and solid phases, respectively, and double Greek superscripts mark the corresponding interfaces. The surface tensions here are thermodynamic, defined as the surface excesses of the grand potential per unit area of the interface. For small droplets, the thermodynamic line tension $\kappa$, defined as the line excess of the grand potential per unit length of the three-phase contact line~\cite{Rusanov-ClassificationLT-ColSurfA-1999, Rusanov-SurfSciRep-2005}, affects the contact angle~\cite{Veselovskii-ZhFizKnim-1936-translit+en, Shcherbakov-ResSurfForces-en-1966, Pethica-RepProgApplChem-1961, *Pethica-JCIS-1977, Rusanov-Deform5-ColloidJUSSR-1977, Rusanov-SurfSciRep-1996, Rusanov-SurfSciRep-2005, Boruvka-JChemPhys-1977}:
\begin{equation}\label{mYe}
\sgmab \! \cos\thetaa = \sgmbg - \sgmag - \frac{\kappa}{\ra}.
\end{equation}
Here, $\theta$ is the contact angle, $r$ is the radius of the three-phase contact line, and the subscript ``a'' refers to an axisymmetric droplet (see Fig.~\ref{fig:droplets}(a)); the line tension $\kappa$ is assumed to be constant. Eq.~\eqref{mYe} is often called the modified Young equation.

By subtracting the modified Young equation~\eqref {mYe} from the classical one~\eqref{cYe} and dividing both parts of the resulting equation by $\sgmab$, one usually arrives at~\cite{Shcherbakov-ResSurfForces-en-1966, Rusanov-Deform5-ColloidJUSSR-1977, Rusanov-SurfSciRep-1996, Rusanov-SurfSciRep-2005}
\begin{equation}\label{mYe-cYe}
\cos\thetabndl - \cos\thetaa = \frac{\kappa}{\sgmab \ra}.
\end{equation}
This equation is commonly employed for contact-angle-based measurements of the line tension in laboratory experiments (see, \mbox{e.\,g.}, \cite{Law-ProgrSurfSci-2017}) and computer modeling (\mbox{e.\,g.}, molecular dynamics simulations~\cite{Bresme-PhysRevLett-1998, Milchev-EurophysLett-2001, Guo-ChinPhysLett-2005, Scocchi-PhysRevE-2011, Weijs-PhysFluids-2011, Barisik-MolSimul-2013, Peng-MolSimul-2014, Isaiev-MolSimul-2015, Khalkhali-JChemPhys-2017, Kanduc-JChemPhys-2017}). According to Eq.~\eqref{mYe-cYe}, the slope of the graph $\cos\thetaa$ vs $1/\ra$ must be equal to $- \kappa/\sgmab$\!. The surface tension $\sgmab$ is typically measured/calculated separately.

However, some authors have observed that Eq.~\eqref{mYe-cYe} is generally not correct~\cite{Tatyanenko-IPHT-2017, Kanduc-PhysRevE-2018, Das-JPhysCondensMatter-2018, Tatyanenko-ColloidJ-2019}, since it assumes the corresponding surface tensions in Eqs.~\eqref{cYe} and \eqref{mYe} to be equal, whereas they correspond to different equilibrium sessile droplet sizes, which implies different values of the chemical potential(s). It has been argued that the adsorption at the solid--liquid interface and its effect on the surface tension $\sgmag$\! affect $\cos\thetaa$ in the same manner and with a comparable magnitude to that observed in experiments~\cite{Ward-PhysRevLett-2008, Ward-AdvColloidInterfaceSci-2010}; thus, the line tension itself is not necessary to explain the dependence of $\cos\thetaa$ on $1/\ra$. As was demonstrated within an interface displacement model, the contributions of the line tension and the adsorption at the solid--gas interface (and its impact on the surface tension $\sgmbg$) to the dependence of $\cos\thetaa$ on $1/\ra$ can be comparable in the presence of a precursor film~\cite{Tatyanenko-IPHT-2017}. Some authors~\cite{Zhang-PhysFluids-2018, Isaiev-MolSimul-2015} assume that only $\sgmab$ depends on the droplet size and Eq.~\eqref{mYe-cYe} is correct, although the slope of the graph $\cos\thetaa$ vs $1/\ra$ yields a quantity different from $- \kappa/\sgmab$\!. Some authors refer to the resulting ``line tension'' as the \emph{apparent} one~\cite{Kanduc-JChemPhys-2017, Kanduc-PhysRevE-2018, Zhang-PhysFluids-2018, Iwamatsu-JAdhesionSciTechnol-2018, Zhao-PhysRevLett-2019, Kubochkin-PhysRevFluids-2021, Klauser-Nanomaterials-2022}. General thermodynamic analysis demonstrates~\cite{Tatyanenko-IPHT-2017, Tatyanenko-ColloidJ-2019} that the value of this slope is influenced by contributions from both the line tension and the adsorptions at all three interfaces (see also \cite{Schimmele-EurPhysJE-2009} for similar arguments).

How does one ``extract'' the line-tension contribution and determine the value of the line tension from the measurements or calculations of $\cos\theta$ vs $1/r$? The thermodynamic analysis~\cite{Tatyanenko-ColloidJ-2019} demonstrates that the equation corresponding to Eq.~\eqref{mYe} for \emph{cylindrical} droplets lacks a $\kappa/r$ term (such droplets are long liquid ``channels'', ``filaments'', or ``ridges'' attached to the substrate surface, as illustrated in Fig.~\ref{fig:droplets}(b); corresponding quantities will be marked with the subscript ``c''). This means that dependence of the contact angle $\thetac$ on the size of a cylindrical droplet (\mbox{e.\,g.}, its half-width $\rc$ on the substrate) is governed mainly by dependences of the surface tensions on the droplet size via the chemical potential(s). This suggests an idea to compare or combine data on the size dependence of $\cos\theta$ for axisymmetric and cylindrical droplets under the same conditions to exclude the surface-tension-related contributions and determine the line tension.

In Sec.~\ref{sec:thermodynamics} we present a thermodynamic analysis of various contributions to droplet-size dependence of the contact angle for both axisymmetric and cylindrical droplets in a system with a single-component fluid and an ideal, smooth, homogeneous solid substrate in the absence of external fields. Based on this analysis, in Sec.~\ref{sec:line_tension-from-contact_angles} we propose a simple technique that allows for the determination of the line tension from a combination of contact-angle size dependences for axisymmetric and cylindrical droplets, similarly to the use of Eq.~\eqref{mYe-cYe}.

Cylindrical droplets are often used in modeling, \mbox{e.\,g.}, in molecular dynamics (MD) simulations~\cite{Dammer-PhysRevLett-2006, Ould-Kaddour-JChemPhys-2011, Scocchi-PhysRevE-2011, Weijs-PhysFluids-2011, Isaiev-MolSimul-2015, Jiang-JChemPhys-2017, Kanduc-Netz-JChemPhys-2017, Kanduc-JChemPhys-2017, Kanduc-PhysRevE-2018, Isaiev-JPhysChemB-2018, Carlson-JPhysChemLett-2024, Burian-SciRep-2024} due to the convenient use of periodic boundary conditions along their longitudinal axes and anticipated weak dependence of the contact angle on the droplet width. However, they are hardly available for experimental measurements due to Plateau--Rayleigh-type instability~\cite{Mechkov-EPL-2007}. Therefore, the technique we propose is primarily oriented towards modeling, especially with methods that do not permit direct calculation of the line excess of the grand potential (\mbox{e.\,g.}, MD simulations). The dimensions of such simulated axisymmetric and cylindrical droplets, \mbox{i.\,e.}, $R$, $r$, and $\theta$, can be ``measured'', and these data can be used to determine the line tension.

\section{\label{sec:thermodynamics}Thermodynamics and Contact Angles}

\subsection{\label{sec:equilibrium-cond}Equilibrium conditions for sessile droplets}

In order to describe the thermodynamic properties of a small sessile droplet on a partially wettable substrate, we consider an open system (that can exchange both energy and matter with its surroundings) of a fixed volume including the droplet itself, the substrate, and the surrounding gas (vapor). The liquid and gaseous phases are assumed to constitute a single-component fluid at given values of chemical potential~$\mu$ and temperature~$T$. These conditions correspond to the grand canonical statistical ensemble. The grand thermodynamic potential of the system can be decomposed into bulk, surface, and line parts:
\begin{equation}\label{GTP-decomposition}
\begin{aligned}
\Omega = &- p^\alpha \Va - p^\beta \Vb + \omega^\gamma V^\gamma \\
&{} + \sgmab \! \Aab + \sgmag \! \Aag + \sgmbg \! \Abg + \kappa L
\end{aligned}
\end{equation}
with $V$ the volume, $A$ the area of the interface, $L$ the length of the three-phase contact line, $p$ the pressure in the bulk phase, and $\omega$ the bulk density of the grand thermodynamic potential (equals $-p$ in fluid phases). Other quantities as well as notations of phases and interfaces are explained above.

Let us consider two types of droplet geometry: (1)~an axisymmetric droplet, shaped as a spherical segment in our ``macroscopic'' thermodynamic description, see Fig.~\ref{fig:droplets}(a), and (2)~a fixed-length piece of a cylindrical droplet, shaped as a segment of a circular cylinder in our ``macroscopic'' thermodynamic description, see Fig.~\ref{fig:droplets}(b). In both geometries, the cross-section of the droplet in our ``macroscopic'' thermodynamic description is a circular segment, thus,
\begin{equation}\label{droplet-geom-rel}
r = R \sin\theta.
\end{equation}
Therefore, the shape and size of the droplet can be described by two independent variables, \mbox{e.\,g.}, $(R, r)$, $(R, \theta)$, or $(r, \theta)$.

In both geometries, the equilibrium conditions for sessile droplets can be obtained by expressing all the volumes, areas, and the length of the three-phase contact line in~\eqref{GTP-decomposition}, and solving one of the following sets of two equations:
\begin{subequations}
\begin{align}
&\partial\Omega(R,r,T,\mu)/\partial R = 0 &\!\!\!\text{and}\!\!\! \quad &\partial\Omega(R,r,T,\mu)/\partial r = 0,\\
&\partial\Omega(R,\theta,T,\mu)/\partial R = 0 &\!\!\!\text{and}\!\!\! \quad &\partial\Omega(R,\theta,T,\mu)/\partial \theta = 0, \label{GTP-R-theta-eq_conds} \\
&\partial\Omega(r,\theta,T,\mu)/\partial r = 0 &\!\!\!\text{and}\!\!\! \quad &\partial\Omega(r,\theta,T,\mu)/\partial \theta = 0,
\end{align}
\end{subequations}
and then taking into account the geometric relation~\eqref{droplet-geom-rel}. Since $T$ and $\mu$ are constant, the pressures and the surface tensions in~\eqref{GTP-decomposition} will be constant at such variations (for $\sgmab$\!, this also requires $\ppartial\sgmab\!\!/\ppartial R = 0$, which is true for the surface of tension chosen as the $\ab$ dividing surface).

The droplet-size-dependent surface tensions $\sgmaba$\! and $\sgmabc$\! for spherical and cylindrical interfaces may differ, although their difference is expected to be very small at the same value of $\mu$ (see Sec.~\ref{sec:pressures-tensions-mu-theta-size-rels}).

Let us chose $R$ and $\theta$ as independent variables to describe the sessile droplet. Introducing nonvariable quantities for the total volume occupied by the fluid $\Vtot \equiv V^\alpha + V^\beta$\! and the total area of the substrate $\Atot \equiv \Aag + \Abg$ and using relation \eqref{droplet-geom-rel}, one can express the volumes, areas, and the contact-line length for an axisymmetric droplet as
\begin{gather*}
\Vaa = \frac{\pi \Ra^3}3 (2 + \cos\thetaa)(1 - \cos\thetaa)^2,
\quad
\La = 2\pi \Ra \sin\thetaa, \\
\Aaba = 2\pi \Ra^2 (1 - \cos\thetaa),
\quad
\Aaga = \pi \Ra^2 \sin^2 \! \thetaa,\\[0.3em]
\Vba = \Vtot - \Vaa,
\quad
\Abga = \Atot - \Aaga.
\end{gather*}
Introducing also $\Dsgmg \equiv\sgmag - \sgmbg$, we can now rewrite decomposition \eqref{GTP-decomposition} of the grand potential as
\begin{align*}
\Omegaa&(\Ra,\thetaa) = - \bigl( p^\alpha \! - p^\beta \bigr) \frac{\pi \Ra^3}3 (2 + \cos\thetaa)(1 - \cos\thetaa)^2  \nonumber \\
&{} + 2 \pi \sgmaba \! \Ra^2 (1 - \cos\thetaa) + \Dsgmg \pi \Ra^2 \sin^2 \!\thetaa + 2\pi \kappaa \Ra \sin\thetaa  \nonumber \\[0.3em]
&{} - p^\beta \Vtot - \sgmbg \! \Atot + \omega^\gamma V^\gamma \end{align*}

We assume the line tension $\kappa = \kappa(T,\mu,r)$ in both geometries. This gives, with use of Eq.~\eqref{droplet-geom-rel},
\begin{align*}
&\biggl(\! \frac{\ppartial\kappa}{\ppartial R} \!\biggr)_{\!\!\theta, T, \mu} \! = \biggl(\! \frac{\ppartial\kappa}{\ppartial r} \!\biggr)_{\!\! T, \mu} \! \biggl(\! \frac{\ppartial r}{\ppartial R} \!\biggr)_{\!\! \theta} = \biggl(\! \frac{\ppartial\kappa}{\ppartial r} \!\biggr)_{\!\! T, \mu} \sin\theta, \\[0.5em]
&\biggl(\! \frac{\ppartial\kappa}{\ppartial \theta} \!\biggr)_{\!\! R, T, \mu} \! = \biggl(\! \frac{\ppartial\kappa}{\ppartial r} \!\biggr)_{\!\! T, \mu} \! \biggl(\! \frac{\ppartial r}{\ppartial \theta} \!\biggr)_{\!\! R} = \biggl(\! \frac{\ppartial\kappa}{\ppartial r} \!\biggr)_{\!\! T, \mu} R \cos\theta.
\end{align*}

Application of the equilibrium conditions \eqref{GTP-R-theta-eq_conds} then yields
\begin{gather*}
\begin{multlined}[t][8.5cm]
- \pi \Ra^2 \bigl( p^\alpha \! - p^\beta \bigr) (2 + \cos\thetaa) (1 - \cos\thetaa)^2 \\
{} + 4 \pi \Ra \sgmaba (1 - \cos\thetaa) + 2 \pi \Ra \Dsgmg \! \sin^2 \! \thetaa \\
{} + 2 \pi \kappaa \sin\thetaa + 2 \pi \Ra \frac{\partial \kappaa}{\partial \ra} \sin^2 \!\thetaa = 0,
\end{multlined}\\[0.5em]
\begin{multlined}
- \pi \Ra^3 \bigl( p^\alpha \! - p^\beta \bigr) \! \sin^3 \!\thetaa + 2 \pi \sgmaba \! \Ra^2 \sin\thetaa + 2 \pi \kappaa \Ra \cos\thetaa \\
{} + 2 \pi \Ra^2 \Dsgmg \! \sin\thetaa \cos\thetaa + 2 \pi \Ra^2 \frac{\partial \kappaa}{\partial \ra} \sin\thetaa \cos\thetaa = 0.
\end{multlined}
\end{gather*}
In order to exclude the pressure difference $p^\alpha - p^\beta$, we multiply the first equation by $(2 \pi \Ra)^{-1} \! \sin^2 \!\thetaa \, (1 - \cos\thetaa)^{-2}$ and then subtract the second equation, multiplied by $(2 + \cos\thetaa)(2 \pi \Ra^2 \sin\thetaa)^{-1}$~\cite{Rusanov-Nanoscale-2021}. After some trigonometric simplifications and the substitution of the result into one of the aforementioned conditions, we arrive at the Laplace equation and the generalized Young equation as equilibrium conditions for an axisymmetric sessile droplet~\cite{Rusanov-Deform5-ColloidJUSSR-1977, Rusanov-SurfSciRep-1996, Rusanov-LTDivSurf-ColSurfA-2004, Rusanov-SurfSciRep-2005, Tatyanenko-IPHT-2017, Tatyanenko-ColloidJ-2019}:
\begin{gather}
p^\alpha - p^\beta = \frac{2\sgmaba}{\Ra}, \label{Lea} \\[0.4em]
\sgmaba \! \cos\thetaa = - \Dsgmg - \frac{\kappaa}{\ra} - \frac{\ppartial \kappaa}{\ppartial \ra} \label{gYea}
\end{gather}

For a cylindrical droplet,
\begin{gather*}
\Vac = \frac{\Lc \Rc^2}{4} (2 \thetac - \sin 2\thetac),\\
\Aabc = \Lc \Rc \thetac,
\quad
\Aagc = \Lc \Rc \sin \thetac,\\[0.5em]
\Vbc = \Vtot - \Vac,
\quad
\Abgc = \Atot - \Aagc,
\end{gather*}
where the length $\Lc$ of the three-phase contact line is independent of $\Rc$ and $\thetac$ (see Fig.~\ref{fig:droplets}(b)) and is assumed to be constant. The contact angle $\thetac$ here is expressed in radians. Decomposition \eqref{GTP-decomposition} of the grand potential then takes the form
\begin{align*}
\!\!\!\! \Omegac&(\Rc,\thetac) = - \bigl( p^\alpha \! - p^\beta \bigr) \frac{\Lc \Rc^2}{4} (2 \thetac - \sin 2\thetac \bigr) + \sgmabc \! \Lc \Rc \thetac  \nonumber \\
&{} + \Dsgmg \Lc \Rc \sin \thetac + \kappac \Lc - p^\beta \Vtot - \sgmbg \! \Atot + \omega^\gamma V^\gamma.
\end{align*}
Application of the equilibrium conditions \eqref{GTP-R-theta-eq_conds} yields
\begin{gather*}
\begin{multlined}[t][8.1cm]
- \bigl( p^\alpha \! - p^\beta \bigr) \Lc \Rc (\thetac - \sin\thetac \cos\thetac) + \sgmabc \! \Lc \thetac \\
{} + \Dsgmg \Lc \sin \thetac + \Lc \frac{\partial \kappac}{\partial \rc} \sin \thetac = 0,
\end{multlined}\\[0.5em]
\begin{multlined}[t][8.1cm]
- \bigl( p^\alpha \! - p^\beta \bigr) \Lc \Rc^2 \sin^2 \!\thetac + \sgmabc \! \Lc \Rc \\
{} + \Dsgmg \Lc \Rc \cos\thetac + \Lc \Rc \frac{\partial \kappac}{\partial \rc} \cos\thetac = 0.
\end{multlined}
\end{gather*}
Here we have employed trigonometric identities $\sin 2\thetac = 2 \sin\thetac \cos\thetac$ and $1 - \cos 2\thetac = 2 \sin^2 \! \thetac$. To derive a more concise equation for $p^\alpha - p^\beta$, we multiply the first equation by $\Rc \cos\thetac$ and subtract the second equation, multiplied by $\sin\thetac$. After some simplifications and the substitution of the result into one of the aforementioned conditions, we arrive at the Laplace equation and the generalized Young equation as equilibrium conditions for a cylindrical sessile droplet, slightly different from Eqs. \eqref{Lea} and~\eqref{gYea}:
\begin{gather}
p^\alpha \! - p^\beta = \frac{\sgmabc}{\Rc}, \label{Lec}\\[0.4em]
\sgmabc \! \cos\thetac = - \Dsgmg - \frac{\ppartial \kappac}{\ppartial \rc}. \label{gYec}
\end{gather}

The absence of the coefficient 2 in the Laplace equation ~\eqref{Lec} for the cylindrical droplet is explained by the fact that its $\ab$ interface is curved along a single direction, whereas the spherical droplet has equal curvatures along two directions. The line-tension terms in the generalized Young equations \eqref{gYea} and \eqref{gYec} originate from variations of the line term $\kappa L$ in~\eqref{GTP-decomposition}. For an axisymmetric droplet, $\La = 2\pi \ra$ and $\ppartial(\!\kappaa \La\!)\!/\ppartial \ra = 2\pi \kappaa + 2\pi \ra \, \ppartial\kappaa/\ppartial\ra$ yield the two terms $\kappaa/\ra + \ppartial\kappaa/\ppartial\ra$ in Eq.~\eqref{gYea}. For a cylindrical droplet, $\Lc$ is fixed, and $\ppartial(\!\kappac \Lc\!)\!/\ppartial \rc = \Lc\, \ppartial\kappac/\ppartial\rc$ yields the only term $\ppartial\kappac\!/\ppartial\rc$ in Eq.~\eqref{gYec}.

For both geometries, we suppose the line tension to depend not only on $T$ and $\mu$, but also on the droplet size. This dependence is typically overlooked, and in this case, Eq.~\eqref{gYea} is transformed into the modified Young equation \eqref{mYe}, whereas Eq.~\eqref{gYec} is transformed into the formally classical Young equation \eqref{cYe} (with surface tensions, however, depending on the chemical potential, \mbox{i.\,e.}, on the equilibrium droplet size, which makes the contact angle $\thetac$ different from $\thetabndl$, as discussed in Sec.~\ref{sec:pressures-tensions-mu-theta-size-rels}). 
We suppose the contact-line radius/half-width $r$ to be the only appropriate size variable the line tension can be considered to depend on, as detailed in appendix~\ref{appx:lt_drop_size_variable}.
The elimination of the size-dependent term $\ppartial\kappa/\ppartial r$ from the generalized Young equations \eqref{gYea} and \eqref{gYec} renders them inexact, if $\kappa$ is assumed to be dependent on the droplet size. It can be seen in a model system if the line tension is calculated directly~\cite{Tatyanenko-IPHT-2017}. Furthermore, the generalized Young equation is the most straightforward approach to determining $\ppartial\kappa/\ppartial r$, provided that $\theta$ and $r$ are found and the surface and line tensions (the latter for axisymmetric droplet only) are given or calculated directly~\cite{Tatyanenko-IPHT-2017}.

\subsection{\label{sec:pressures-tensions-mu-theta-size-rels}Size dependence of contact angle and relations between pressures, surface tensions, droplet size, and chemical potential}

A comparison of the generalized Young equation \eqref{gYea} or \eqref{gYec} with the classical Young equation \eqref{cYe} must take into account the fact that these two equations are written for droplets corresponding to \emph{different} values of the chemical potential $\mu$. The classical Young equation is applicable to a macroscopically large (strictly speaking, infinite) drop, \mbox{i.\,e.}, the liquid--gas bulk phase coexistence at the binodal. Let us mark the values of the quantities corresponding to the binodal with the subscript ``$\infty$''. The surface tensions at the binodal then should be denoted as $\sgmabbndl$\!, $\sgmagbndl$\!, $\sgmbgbndl$, and thus, $\Dsgmgbndl \equiv \sgmagbndl \! - \sgmbgbndl$. Consequently, the classical Young equation \eqref{cYe} can be rewritten as
\begin{equation}\label{cYem}
\sgmabbndl \! \cos \thetabndl = - \Dsgmgbndl.
\end{equation}

By subtracting the generalized Young equations \eqref{gYea} and \eqref{gYec} from Eq.~\eqref{cYem}, we obtain two similar equations: one for an axisymmetric droplet
\begin{gather}
\sgmabbndl \! \cos\thetabndl - \sgmaba \! \cos\thetaa = \delta\Dsgmg + \frac{\kappaa}{\ra} + \frac{\ppartial \kappaa}{\ppartial r} \label{cYem-gYea} \\[-0.2em]
\intertext{and the other for a cylindrical droplet}
\sgmabbndl \! \cos\thetabndl - \sgmabc \! \cos\thetac = \delta\Dsgmg + \frac{\ppartial \kappac}{\ppartial \rc}. \label{cYem-gYec}
\end{gather}
Here $\delta\Dsgmg \equiv \Dsgmg - \Dsgmgbndl$. To make these equations more similar to Eq.~\eqref{mYe-cYe}, let us introduce $\delta\sgmabac \equiv \sgmabac - \sgmabbndl$. Then, Eqs.~\eqref{cYem-gYea} and \eqref{cYem-gYec} can be rewritten as
\begin{align}
&\!\!\!\!\! \sgmabbndl \! ( \cos \thetabndl - \cos\thetaa ) = \delta\Dsgmg \! + \delta\sgmaba\!\!\cos\thetaa + \frac{\kappaa}{\ra} \! + \! \frac{\ppartial \kappaa}{\ppartial \ra}, \! \label{cYem-gYea-2} \\[0.5em]
&\!\!\!\!\! \sgmabbndl \! ( \cos\thetabndl - \cos\thetac) = \delta\Dsgmg \! + \delta\sgmabc\!\!\cos\thetac + \frac{\ppartial \kappac}{\ppartial \rc}. \label{cYem-gYec-2}
\end{align}

It can be observed that Eq.~\eqref{cYem-gYea-2}, the exact counterpart of Eq.~\eqref{mYe-cYe} for the difference $\cos\thetabndl - \cos\theta$, contains four terms for an axisymmetric droplet (instead of one) and three terms for a cylindrical droplet (instead of none). The question thus arises as to how these terms depend on the droplet size (the radius/half-width of the droplet on the substrate).

For the $\ab$ liquid--gas interface, the isothermal Gibbs adsorption equations can be written as
\begin{equation}\label{Gae_ab}
\dif\sgmabac = - \Gabac \! \dif\mu,
\end{equation}
where $\Gab$ is the adsorption (the surface excess of matter per unit surface area) at the $\ab$ interface for corresponding geometry. As the surface of tension is chosen as the dividing surface in both geometries, these equations do not contain $(\ppartial\sgmab\!\!/\ppartial R) \dif R$ terms.

For each of the solid--fluid interfaces, similar equations can be written. In general, these equations are more complicated (see, \mbox{e.\,g.}, Sec.~6.3 in \cite{Rusanov-SurfSciRep-1996} and Sec.~3.4.3 in \cite{Rusanov-SurfSciRep-2005}). However, in certain simple cases (\mbox{e.\,g.}, for nondeformable substrate and such a choice of the solid--fluid dividing surfaces that the amounts of matter of the immobile component of solid do not change in the solid phase at any change of $\mu$), they take~\cite{Tatyanenko-IPHT-2017, Tatyanenko-ColloidJ-2019} the same simple form as~\eqref{Gae_ab}
\begin{equation}\label{Gae_Dg}
\dif\Dsgmg \equiv \dif \bigl( \sgmag - \sgmbg \bigr) = \bigl( \Gbg - \Gag \bigr) \dif\mu.
\end{equation}
By integrating Eqs.~\eqref{Gae_ab} and \eqref{Gae_Dg} from $\mubndl$ to $\mu$, we obtain, introducing $\delta\mu\equiv\mu - \mubndl$,
\begin{gather}
\delta\sgmabac = - \! \int_{\mubndl}^{\mu} \Gabac\!(\mu') \dif\mu' \simeq - \Gabbndl \delta\mu,  \label{sgmab_mu} \\[0.5em]
\delta\Dsgmg = \int_{\mubndl}^{\mu} \bigl( \Gbg - \Gag \bigr) \dif\mu' \simeq \bigl( \Gbgbndl - \Gagbndl \bigr) \delta\mu. \label{dDsgm_mu}
\end{gather}
Since $\Gaba\!(\mubndl) = \Gabc\!(\mubndl) = \Gabbndl$\!, the asymptotic expression in~\eqref{sgmab_mu} is identical for spherical and cylindrical surfaces at the same value of $\mu$, but the exact integral is generally not the same if $\Gaba\!(\mu) \neq \Gabc\!(\mu)$. Thus, $\sgmaba$\! and $\sgmabc$\! generally differ, but at least in the second order in $\delta\mu$,
\begin{equation}\label{sgmab_a_c-difference}
\sgmaba \! - \sgmabc = \delta\sgmaba \! - \delta\sgmabc = \bigO\bigl(\!(\delta\mu)^{2\,}\!\bigr).
\end{equation}

The last asymptotic expressions in Eqs.~\eqref{sgmab_mu} and \eqref{dDsgm_mu} are valid for small values of $\delta\mu$. What is the relationship between this quantity and the droplet size? This can be estimated using the Laplace equation \eqref{Lea} or \eqref{Lec} for the axisymmetric or cylindrical droplet, respectively. By applying the isothermal Gibbs--Duhem equations $\dif p = n \dif \mu$ to the liquid and gaseous phases, we obtain 
\begin{equation}\label{G-D}
p^\alpha \! - p^\beta \! \simeq \nabndl \bigl( 1 + \chiabndl \delta \! p^\alpha \bigr) \delta\mu \simeq \nabndl\delta\mu + \bigO\bigl(\!(\delta\mu)^{2\,} \! \bigr)\!
\end{equation}
with $\na$ the number density of molecules and $\chia$ the isothermal compressibility in the liquid phase, $\delta \! p^\alpha \equiv p^\alpha \! - \pabndl$. Here we have neglected the number density of molecules in the gas ($\nb \ll \na$). We will further neglect the compressibility of the liquid since the resulting correction is only of the second order in $\delta\mu$. Using the Laplace equations \eqref{Lea} and \eqref{Lec}, along with the geometric relation \eqref{droplet-geom-rel}, we obtain
\begin{gather}
\frac{2\sgmaba}{\Ra} = \frac{2\sgmaba \!\! \sin\thetaa}{\ra} \simeq \nabndl\delta\mu, \label{R-r-mu-rel-a}\\[0.5em]
\frac{\sgmabc}{\Rc} = \frac{\sgmabc \!\! \sin\thetac}{\rc} \simeq \nabndl\delta\mu. \label{R-r-mu-rel-c}
\end{gather}
It can now be seen that $\delta\mu \propto 1/R = \bigO(1/r)$, \mbox{i.\,e.}, small values of $\delta\mu$ correspond to large droplets, and $\delta\mu \to 0$ expectedly corresponds to $R \to \infty$ and $r \to \infty$.

Accordingly, the correction terms in Eqs.~\eqref{cYem-gYea-2} and \eqref{cYem-gYec-2} are given by
\begin{gather}
\delta\Dsgmg \simeq \frac{2\sgmaba \! \sin\thetaa}{\nabndl \ra} \bigl( \Gbgbndl - \Gagbndl \bigr) \! = \bigO(\delta\mu) = \bigO(1/r), \label{dDsgr-a}\\[0.4em]
\delta\sgmaba \simeq - \frac{2\sgmaba \Gabbndl \! \sin\thetaa}{\nabndl \ra} = \bigO(\delta\mu) = \bigO(1/r), \label{dsgmab-a}\\[0.4em]
\kappaa/\ra \simeq \kappabndl/\ra = \bigO(\delta\mu) = \bigO(1/r), \label{kappaa_ra-asymt}
\intertext{where $\kappabndl = \kappaa$ at $\mu = \mubndl$ ($\ra \to \infty$) is the line tension of the macroscopic (infinite) droplet, and}
\delta\Dsgmg \simeq \frac{\sgmabc \! \sin\thetac}{\nabndl \rc} \bigl( \Gbgbndl - \Gagbndl \bigr) = \bigO(\delta\mu) = \bigO(1/r), \label{dDsgr-c} \\[0.4em]
\delta\sgmabc \simeq - \frac{\sgmabc \Gabbndl \! \sin\thetac}{\nabndl \rc} = \bigO(\delta\mu) = \bigO(1/r). \label{dsgmab-c}
\end{gather}

With regard to the terms $\ppartial\kappa/\ppartial r$, it can be shown~\cite{Rusanov-LTDivSurf-ColSurfA-2004, Rusanov-SurfSciRep-2005, Tatyanenko-IPHT-2017, Tatyanenko-ColloidJ-2019} that they are $\bigO \bigl(\!(\delta\mu)^{2\,}\!	\bigr) = \bigO \bigl( 1/r^2 \bigr)$. To demonstrate this, the line adsorption equation~\cite{Chen-ColSurfA-2000} (a line counterpart of the Gibbs adsorption equation and the Gibbs--Duhem equation) can be considered~\cite{Rusanov-LTDivSurf-ColSurfA-2004, Rusanov-SurfSciRep-2005, Tatyanenko-ColloidJ-2019} at a constant temperature:
\begin{equation*}
\dif \kappa = - \Lambda \dif \mu + ( \ppartial \kappa / \ppartial r )_{\!\mu,T} \dif r
\end{equation*}
with $\Lambda$ the line excess of the fluid matter (the number of molecules of the fluid) per unit length of the three-phase contact line (a line counterpart of the adsorption~$\Gamma$) in the corresponding geometry. Assuming then
\begin{gather}
\kappa = \kappabndl + \bigO(\delta\mu),
\qquad
\Lambda = \Lambdabndl + \bigO(\delta\mu) \label{kappa-Lambda-asympt} \\
\intertext{along the equilibrium curve $r(\mu)$ with the surface of tension chosen as the liquid--gas dividing surface at each value of $\mu$, we arrive at the estimation}
\!\!\!\!\!\! \biggl(\!\! \frac{\ppartial \kappa}{\ppartial r} \!\! \biggr)_{\!\!\!\mu,T} \!\! = \frac{\dif \kappa}{\dif r} + \Lambda	 \frac{\dif \mu}{\dif r} = \left[ \frac{\dif \kappa}{\dif \mu} + \Lambda \right] \!\!\! \biggl( \!\! \frac{\dif r}{\dif \mu} \!\! \biggr)^{\!\!-1} \!\!\!\! = \bigO\bigl(\!(\delta\mu)^2\bigr).  \label{pdkappa_r-estimation}
\end{gather}
The factor in square brackets is $\bigO(1)$ since \eqref{kappa-Lambda-asympt}, while $\dif r/\dif \mu = \dif (R \sin\theta)/\dif \mu \simeq (\dif R/\dif \mu) \sin\thetabndl = \bigO \bigl(\! (\delta\mu)^{\!-2}\bigr)$. This estimation is supported, in the case of $\ppartial\kappaa/\ppartial\ra$, by direct calculations within an interface displacement model~\cite{Tatyanenko-IPHT-2017}.

This consideration is somewhat simplified; it does not take into account the influence of the stress tensor distribution in the solid substrate on the line tension. We assume that the estimation $\ppartial \kappa / \ppartial r = \bigO\bigl(\!(\delta\mu)^2\bigr) = \bigO\bigl(1/r^2\bigr)$ remains valid if the corresponding contribution is taken into account.

For sufficiently large droplets, $\sin\theta$ can be approximated with $\sin\thetabndl$, with a relative error $\bigO(\delta\mu) = \bigO(1/r)$. This makes Eqs.~\eqref{dDsgr-a}, \eqref{dsgmab-a}, \eqref{dDsgr-c}, and \eqref{dsgmab-c} valid with $\sin\thetaac$ replaced by $\sin\thetabndl$. The same approximation can be applied to $\cos\theta$ in the correction terms $\delta\sgmab\!\cos\theta$ in Eqs.~\eqref{cYem-gYea-2} and~\eqref{cYem-gYec-2}.

In light of these considerations, we can now proceed to derive the linear approximations in $1/r$ for Eqs.~\eqref{cYem-gYea-2} and \eqref{cYem-gYec-2}:
\begin{gather}
\cos\thetabndl - \cos\thetaa \simeq \frac{\kappaappa}{\sgmabbndl \! \ra}, \label{cYem-gYea-1st-order}\\[0.4em]
\cos\thetabndl - \cos\thetac \simeq \frac{\kappaappc}{\sgmabbndl \! \rc}, \label{cYem-gYec-1st-order}\end{gather}
\vspace{-1.6em}
\noindent
where\\
\begin{align}
&\kappaappa \equiv \kappabndl + \frac{2 \sgmabbndl \! \sin\thetabndl}{\nabndl} \bigl( \Gbgbndl - \Gagbndl - \Gabbndl\! \cos\thetabndl \bigr) \label{applt-a}\\
\intertext{and}
&\kappaappc \equiv \frac{\sgmabbndl \! \sin\thetabndl}{\nabndl} \bigl( \Gbgbndl - \Gagbndl - \Gabbndl\! \cos\thetabndl \bigr) \label{applt-c}
\end{align}
are the quantities sometimes referred to as the \emph{apparent line tensions}~\cite{Kanduc-JChemPhys-2017, Kanduc-PhysRevE-2018, Zhang-PhysFluids-2018, Iwamatsu-JAdhesionSciTechnol-2018, Zhao-PhysRevLett-2019, Kubochkin-PhysRevFluids-2021, Klauser-Nanomaterials-2022} of axisymmetric and cylindrical droplets, respectively (see below).

Equations~\eqref{cYem-gYea-1st-order} and \eqref{applt-a} demonstrate that the dependence of $\cos\thetaa$ on $1/\ra$, even in the first order of approximation, is determined not only by the line tension, but also by the adsorptions at the three interfaces. Indeed, Eq.~\eqref{cYem-gYea-1st-order} should be treated as Eq.~\eqref{mYe-cYe} typically is. This implies that the slope of the graph $\cos\thetaa$ vs $1/\ra$ at $1/\ra \to 0$ does not directly yield the line tension; rather, it represents a combination of the line tension, the surface tension of the droplet free surface, adsorptions at the three interfaces, and the macroscopic contact angle. Some authors use the term ``apparent line tension'' to refer to the calculated value of the line tension from the $\cos\theta(1/r)$ dependence at $1/r \to 0$, as if Eq.~\eqref{mYe-cYe} were the correct linear approximation. Then Eq.~\eqref{applt-a} yields the value of the apparent line tension of axisymmetric droplets.

In previous studies, the impact of individual adsorptions on the contact angle has been considered and estimated.

It is likely that the most substantial contribution can be made by the solid--gas adsorption $\Gbg$ for a droplet formed on the top of a precursor film. The formation of such a film can, \mbox{e.\,g.}, precede dropwise condensation on a partially wettable substrate~\cite{Popescu-JPhysCondensMatter-2012}. Such precursor films can be as thick as nanometers (\mbox{e.\,g.}, 3--11\,nm for alcohols and water on a glass~\cite{DerjaguinZorin-ZhFizKhim-1955-translit+en, *DerjaguinZorin-ProgSurfSci-1992} and 11--12\,nm for water on fused quartz~\cite{Hall-JPhysChem-1970} for a film in equilibrium with saturated vapors). In such a case, if the contact angles of sessile droplets are measured at the substrate surface (not at the top of the precursor film), this film should be considered as an equilibrium state of the solid substrate and treated as an adsorption film resulting in a high adsorption $\Gbgbndl \approx \fbndl \nabndl$ (with $\fbndl$ the precursor film thickness at equilibrium with an infinitely large droplet). Therefore, its contribution to the apparent line tension $\kappaappa$ can be estimated as $2 \sgmabbndl \!\! \fbndl \sin\thetaa \lesssim 10^{-10}$--$10^{-9}$\,N.

The line tension and the solid--gas adsorption contribution have been directly calculated within an interface displacement model with a model short-range interface potential~\cite{Tatyanenko-IPHT-2017}. The results indicated that they can be comparable in magnitude, with the solid--gas adsorption contribution being even several times more in magnitude than the line tension $\kappabndl$. It should be noted that other adsorptions are zero within this model.

The impact of the adsorption $\Gag$ at the solid--liquid interface on the contact angle was examined by Ward and Wu~\cite{Ward-PhysRevLett-2008} in the context of a discussion of $\cos\thetaa(1/\ra)$ measurements for dodecane sessile droplets on silanized silicon wafers~\cite{Checco-PhysRevLett-2003}. The adsorption $\Gbg$ at the solid--gas interface was deemed negligible compared to $\Gag$ due to the low volatility of dodecane~\cite{Ward-PhysRevLett-2008}. The calculations demonstrated that the effect of the adsorption at the solid--liquid interface was sufficient to explain the observed droplet-size-dependence of the contact angle, even without the line tension. These arguments were subsequently developed in~\cite{Ward-AdvColloidInterfaceSci-2010}.

It is quite common to consider solely the size dependence of the liquid--gas surface tension, which corresponds to taking into account the adsorption $\Gabbndl$\!. The Tolman equation $\sgmab \simeq \sgmabbndl \bigl( 1 - 2 \TolmanLength / \Ra \bigr)$ with the Tolman length~\cite{Tolman-JChemPhys-1949} $\TolmanLength \approx \Gabbndl\!\!\!/\nabndl$ is the most commonly employed and provides the contribution to the apparent line tension $\kappaappa$ equal to $-2 \sgmabbndl \TolmanLength \sin\thetaa \cos\thetaa$. With typical estimates of $\TolmanLength \sim 10^{-10}$\,m and $\sgmabbndl \sim 10^{-2}$\,N/m, this contribution can be estimated as $10^{-12}$\,N, which is smaller or comparable with the values of the apparent line tension $\kappaappa$ observed in simulations~\cite{Kanduc-PhysRevE-2018} and some experiments~\cite{Checco-PhysRevLett-2003}.

Of greater interest is the fact that Eqs.~\eqref{cYem-gYec-1st-order} and \eqref{applt-c} demonstrate that, for cylindrical droplets, $\cos\thetac$ also depends on $1/\rc$ in the first order in $1/\rc$. This has been observed in simulations~\cite{Kanduc-JChemPhys-2017, Kanduc-PhysRevE-2018} and attributed to the size dependence of the surface tension $\sgmab$\! expressed using the Tolman length~\cite{Tolman-JChemPhys-1949} $\TolmanLength \approx \Gabbndl\!\!\!/\nabndl$. As it is seen from Eq.~\eqref{applt-c}, the dependence of other surface tensions on the droplet size via the chemical potential can result in their contributions to the apparent line tension of cylindrical droplets.

\section{\label{sec:line_tension-from-contact_angles}Line tension from dual-geometry contact-angle dependences on droplet size}

One can easily notice that the apparent line tension \eqref{applt-c} of cylindrical droplets, which is proportional to the slope of the $\cos\thetac(1/\rc)$ graph at $1/\rc \to 0$, is exactly half of the second, adsorption-related term in the apparent line tension \eqref{applt-a} of axisymmetric droplets, which is proportional to the slope of the $\cos\thetaa(1/\ra)$ graph at $1/\ra \to 0$. Thus, by combining data for the dependences $\cos\thetaa(1/\ra)$ and $\cos\thetac(1/\rc)$, one can determine the line tension of a macroscopic droplet as
\begin{equation}\label{lt_from_applt}
\kappabndl = \kappaappa - 2\kappaappc.
\end{equation}

\begin{figure}[tb]
\includegraphics[width=8.6cm]{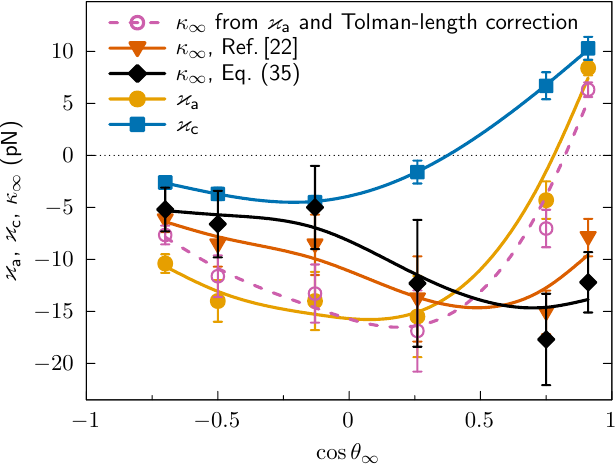}
\caption{\label{fig:lt-kanduc-dg} Apparent line tensions of axisymmetric ($\kappaappa$) and cylindrical ($\kappaappc$) MD-simulated droplets~\cite{Kanduc-JChemPhys-2017, Kanduc-PhysRevE-2018} (data taken from Fig.~2(c) in \cite{Kanduc-PhysRevE-2018}) and corresponding line tension $\kappabndl$ of macroscopic droplets calculated with Eq.~\eqref{lt_from_applt} and by the authors of \cite{Kanduc-PhysRevE-2018} using additional corrections (data taken from Fig.~3 therein) vs the macroscopic contact-angle cosine (tuned by surface polarity). The hollow points and the dashed line show estimations of $\kappabndl$ from the values of $\kappaappa$ and the Tolman-length correction with Eq.~\eqref{applt-a}, corresponding to the $\ab$ (liquid--gas) adsorption correction. An error for $\kappabndl$ from Eq.~\eqref{lt_from_applt} is estimated as the sum of an error for $\kappaappa$ and a doubled error for $\kappaappc$. The lines are weighted cubic splines, plotted using gnuplot's~\cite{gnuplot-6_0-manual} \texttt{smooth acsplines} method with the weights $w/\Delta$, where $\Delta$ is the absolute error of the corresponding value, and $w = 500$\,pN.}
\end{figure}

By employing this relation, we have calculated the line tension $\kappabndl$ of a macroscopic droplet from the apparent line tensions $\kappaappa$ and $\kappaappc$ of sessile droplets simulated by Kandu\v{c}\textit{et~al.}~\cite{Kanduc-JChemPhys-2017, Kanduc-PhysRevE-2018} using data from Fig.~2(c) in \cite{Kanduc-PhysRevE-2018}. As illustrated in Fig.~\ref{fig:lt-kanduc-dg}, the calculated values of $\kappabndl$ using Eq.~\eqref{lt_from_applt} are in good agreement with those obtained in \cite{Kanduc-PhysRevE-2018} by taking into account the Tolman-length and ``line-tension-stiffness'' $\ppartial\kappa/\ppartial\theta$ corrections. In~\cite{Kanduc-JChemPhys-2017, Kanduc-PhysRevE-2018}, the solid--liquid and solid--gas surface tensions are assumed to be constant. However, the equilibrium equations for the contact angles $\thetaac$ derived there are different from those presented here. The discrepancies between these works and the present study in the initial assumptions and the resulting equations will be discussed in Sec.~\ref{sec:results-discussion}.

Nevertheless, the Tolman-length correction can be accounted for through the use of Eq.~\eqref{applt-a}. If we consider solely the correction to $\sgmabbndl \! = 55$\,mN/m with the Tolman length $\TolmanLength \approx \Gabbndl\!\!\!/\nabndl = -0.05$\,nm as determined from simulations~\cite{Kanduc-JChemPhys-2017}, this leads to $\sgmabbndl \! \TolmanLength = -2.75$\,pN, and calculate $\kappabndl$ using Eq.~\eqref{applt-a} with $\Gagbndl \! = \Gbgbndl \! = 0$, the resulting value of $\kappabndl$ will be quite close to $\kappaappa$ (shown with hollow points and a dashed line in Fig.~\ref{fig:lt-kanduc-dg}). Therefore, it can be concluded that the Tolman-length correction alone underestimates the adsorption-related contributions to the apparent line tension of axisymmetric droplets.

It is clearly seen that the values of $\kappaappa$ and $\kappaappc$ are comparable in these simulations, which makes $\kappaappa$ visibly different from $\kappabndl$ as calculated in~\cite{Kanduc-PhysRevE-2018} and with the use of Eq.~\eqref{lt_from_applt}.

Equations~\eqref{cYem-gYea-1st-order}--\eqref{applt-c} suggest another idea. In the context of dual-geometry sessile droplet measurements, there is no need to employ the classical Young equation at all. By subtracting the generalized Young equation for axisymmetric droplets \eqref{gYea} from the one for cylindrical droplets \eqref{gYec}, we arrive at the following equation:
\begin{equation}\label{cYea-gYec}
\sgmabc \! \cos \thetac - \sgmaba \! \cos\thetaa = \frac{\kappaa}{\ra} + \frac{\ppartial \kappaa}{\ppartial \ra} - \frac{\ppartial \kappac}{\ppartial \rc}.
\end{equation}
In consideration of Eqs.~\eqref{dsgmab-a} and \eqref{dsgmab-c} with estimation \eqref{pdkappa_r-estimation}, the first-order approximation to the combined equation \eqref{cYea-gYec} can be expressed as follows:
\begin{equation}\label{cYea-gYec-1st-order}
\cos\thetac - \cos\thetaa \simeq \frac{\kappabndl}{\sgmabbndl \! \ra}.
\end{equation}
The dependence of $\kappaa$ on $\mu$ and $\ra$, as well as the dependence of $\sgmab$\! on $\mu$, are not taken into account since they result in corrections of order $\bigO \bigl( 1/\ra^2 \bigr)$.

In order to exclude the terms affected by changes in surface tension, it is necessary to take all the quantities in Eqs.~\eqref{cYea-gYec} and \eqref{cYea-gYec-1st-order} at the same values of the chemical potential and temperature. The chemical potential is often unknown in simulations or measurements. However, as can be seen from Eqs.~\eqref{R-r-mu-rel-a} and \eqref{R-r-mu-rel-c}, and taking into account Eq.~\eqref{sgmab_a_c-difference}, the same value of $\delta\mu$ corresponds to the relation $\Rc \simeq \Ra/2$ or, in the same order of $\delta\mu$, to $\rc \simeq \ra/2$.

Thus, to determine the line tension from the measured (in the experiment or modeling) dependences of the $\cos\theta$ on $1/r$ for both axisymmetric and cylindrical droplets under the same conditions, it is sufficient to calculate the difference
\begin{equation}\label{diffs_cos_theta-ac}
\! \cos\thetac\Big|_{\Rc = \Ra/2} \! - \cos\thetaa\Big|_{\Ra} \!\quad\! \text{or} \quad \cos\thetac\Big|_{\rc = \ra/2} \! - \cos\thetaa\Big|_{\ra}
\end{equation}
and then take a linear fit of this quantity vs $1/\ra$ at small values of $1/\ra$. The slope of the resulting graph will yield the value of $\kappabndl/\sgmabbndl$:
\begin{gather}
\cos\thetac \Big|_{\Rc = \Ra/2} \! - \cos\thetaa \Big|_{\Ra} \simeq \frac{\kappabndl}{\sgmabbndl \! \ra}, \label{diff_cos_theta_R-linear-ra}\\[0.7em]
\cos\thetac \Big|_{\rc = \ra/2} \! - \cos\thetaa \Big|_{\ra} \simeq \frac{\kappabndl}{\sgmabbndl \! \ra}. \label{diff_cos_theta_r-linear-ra}
\end{gather}

An additional advantage of this technique is that the graph of the difference $\cos\thetac - \cos\thetaa$ vs $1/\ra$ must start from the known point (0 at $1/\ra = 0$). Therefore, the slope of the linear graph remains the only fitting parameter. When plotting $\cos\thetaa(1/\ra)$ and $\cos\thetac(1/\rc)$ independently, each of them must be approximated by a straight line starting from an unknown point $\cos\thetabndl(0)$. This point must be the same for both axisymmetric and cylindrical droplets. In principle, this should be taken into account to reduce the number of fitting parameters from 4 to 3. In contrast to experimental measurements, the computational expense of simulating droplet sizes with macroscopic contact angles can be significant. However, the macroscopic contact angle $\thetabndl$ can be found, \mbox{e.\,g.}, by employing the classical Young equation \eqref{cYem} and calculating $\sgmabbndl$\! and $\Dsgmgbndl$ in three simulations of planar interfaces~\cite{Carlson-JPhysChemLett-2024}.

The use of dual-geometry techniques to determine the line tension may offer another potential advantage, as suggested by experiments with nanosized droplets. For nanosized droplets, the graph $\cos\thetaa$ vs $1/\ra$ can be fairly linear; however, such a linear fit applicable to small droplets may yield a false value of $\cos\thetabndl$ at extrapolation down to $1/\ra \to 0$ (see Fig.~5 in~\cite{Heim-Langmuir-2013}) and the slope that does not correspond to asymptotic Eq.~\eqref{cYem-gYea-1st-order}. In many simulations, the droplets are also nanosized, and thus the same issue may arise with simulated droplets, both axisymmetric and cylindrical, resulting in miscalculation of $\kappaappa$ and $\kappaappc$. The application of Eq.~\eqref{lt_from_applt} would ultimately yield an incorrect value of $\kappabndl$. However, not only would the slope be incorrect in such a situation, but the extrapolated values of $\cos\thetabndl$ would also be different for $\cos\thetaa$ vs $1/\ra$ and $\cos\thetac$ vs $1/\rc$, indicating an inconsistency. The technique we proposed here would reveal such a situation as a set of points on the graph of quantity \eqref{diffs_cos_theta-ac} vs $1/\ra$ that do not fit a straight line originating at the point $(0,0)$. Both these indicators imply that the droplets are not sufficiently large for the asymptotic equations, Eqs.~\eqref{cYem-gYea-1st-order}, \eqref{cYem-gYec-1st-order}, and \eqref{cYea-gYec-1st-order}, to be applicable, thereby precluding the determination of the contact angle $\thetabndl$ and the line tension $\kappabndl$ of macroscopic droplets.

\section{\label{sec:results-discussion}Overview of Results and discussion}

The thermodynamic analysis performed demonstrates that droplet-size corrections to the contact-angle cosine for axisymmetric sessile droplets typically attributed to line tension alone, are generally determined by line tension $\kappaa$, its dependence on droplet size (directly on the droplet radius $\ra$ on the surface and via the chemical potential), and also by dependences of all three surface tensions on droplet size (via the chemical potential). The same applies to the corrections to the contact-angle cosine for cylindrical sessile droplets, except for the line-tension term $\kappac/\rc$ that is absent in the generalized Young equation~\eqref{gYec} for cylindrical droplets.

In the first order in $1/\ra$ \eqref{cYem-gYea-1st-order}, this gives the slope of $\cos\thetaa$ vs $1/\ra$ to be proportional not to the line tension $\kappabndl$ of macroscopic droplets (as it is widely believed according to Eq.~\eqref{mYe-cYe}) but rather to the so-called ``apparent line tension'' $\kappaappa$, a sum of this line tension $\kappabndl$ and a combination of adsorptions at the three interfaces, the number density of molecules in the liquid phase, and the macroscopic contact angle \eqref{applt-a}.

For cylindrical droplets, the contact angle is found to depend on the droplet half-width $\rc$ on the substrate already in the first order in $1/\rc$ \eqref{cYem-gYec-1st-order}, thereby yielding the apparent line tension $\kappaappc$. The latter is determined by the same combination of the adsorptions, the number density of molecules in the liquid phase, and the macroscopic contact angle \eqref{applt-c} but lacks the numerical coefficient 2 compared to the axisymmetric case \eqref{applt-a}. This gives a possibility for determining the line tension $\kappabndl$ from $\kappaappa$ and $\kappaappc$ found in dual-geometry measurements (for droplets simulated at the same conditions, \mbox{i.\,e.}, under the same temperature with the same interaction potentials, etc.) as~\eqref{lt_from_applt}. This was illustrated with calculations of $\kappabndl$ from $\kappaappa$ and $\kappaappc$ of MD-simulated droplets~\cite{Kanduc-JChemPhys-2017, Kanduc-PhysRevE-2018} (see Fig.~\ref{fig:lt-kanduc-dg}).

We have also proposed a technique for dual-geometry measurements based on the calculation of the slope of the difference $\cos\thetac - \cos\thetaa$ \emph{at the same value of the chemical potential} vs $1/\ra$. It can be implemented as a calculation of this difference \eqref{diffs_cos_theta-ac} at $\Rc = \Ra/2$ or $\rc = \ra/2$. The slope will equal $\kappabndl/\sgmabbndl$, with $\sgmabbndl$ the surface tension of the planar liquid--gas interface. The advantages of this technique are discussed in Sec.~\ref{sec:line_tension-from-contact_angles}.

The Laplace and the generalized Young equations in both geometries, as well as the Gibbs adsorption equation \eqref{Gae_ab} for the liquid--gas interface, were obtained and considered for the surface of tension as the liquid--gas dividing surface. At an arbitrary choice of this dividing surface, all these equations become more complicated~\cite{Rusanov-LTDivSurf-ColSurfA-2004, Schimmele-JChemPhys-2007}. Even though most researchers use these equations in their simple forms, the real choice of the dividing surface is often different or even not specified explicitly. In simulations, this can be done, \mbox{e.\,g.}, by employing a boundary detection algorithm, and different algorithms can yield~\cite{Polovinkin-ColloidsSurfA-2024} visibly different dependences of $\cos\thetaa$ vs $1/\ra$. As demonstrated in~\cite{Burian-SciRep-2024}, the dependence of $\cos\thetac$ on $\Rc$ is visibly different for the surface of tension and the equimolecular liquid--gas dividing surface in the same simulated argon cylindrical nanodroplets. Thus, the choice of the liquid--gas dividing surface can be considered as an additional source of errors that need to be estimated.

As for the solid--fluid dividing surfaces, the analog of the Gibbs adsorption equation~\eqref{Gae_Dg} has been employed to express the difference $\Dsgmg \equiv \sgmag - \sgmbg$, implying certain conditions that are not necessarily met at an arbitrary choice of these dividing surfaces. However, if the same choice is made in both geometries, the quantities $\delta\Dsgmg \equiv \Dsgmg - \Dsgmgbndl$ will cancel out, both with use of Eq.~\eqref{lt_from_applt} and in the proposed technique where Eq.~\eqref{cYea-gYec} will remain valid. In contrast, the line tensions $\kappaa$ and $\kappac$ (and their macroscopic value $\kappabndl$) do generally depend on the choice of the solid--fluid dividing surfaces. For any specific choice, the dual-geometry approach must work, giving the value of $\kappabndl$ corresponding to the chosen dividing surfaces.

The idea of employing dual-geometry measurements to determine the line tension has in fact been exploited by Kandu\v{c} \textit{et~al.}~\cite{Kanduc-PhysRevE-2018}. However, there are substantial differences from our findings in details and results. In that work, the line tension is supposed to depend on the contact angle $\theta$ but not the radius/half-width $r$ of the droplet on the substrate. We consider the contact angle $\theta$ to be an inappropriate variable to characterize the size dependence of the line tension (see appendix~\ref{appx:lt_drop_size_variable} for further details). Another difference is that only the liquid--gas surface tension $\sgmab$ is considered to depend on the droplet size (specifically, $R$) in~\cite{Kanduc-JChemPhys-2017, Kanduc-PhysRevE-2018}, while, in our consideration, all three surface tensions are generally dependent on the droplet size. The Tolman length $\TolmanLength \approx \Gabbndl\!\!\!/\nabndl $ was obtained independently in~\cite{Kanduc-JChemPhys-2017} through modeling of a free cylindrical droplet. In our results, all necessary information to account for the effects of adsorptions (the droplet-size dependence of the surface tensions) is contained in the measured dependence $\cos\thetac(1/\rc)$.

The equilibrium equations for the contact angles derived in~\cite{Kanduc-JChemPhys-2017, Kanduc-PhysRevE-2018} are also different from ours, even at constant $\sgmag$, $\sgmbg$, and $\kappa$. This discrepancy can be attributed primarily to the omission of the bulk term in the free-energy-based work of the droplet formation, for which then a stationary point was found under condition of constant droplet volume~\cite{Kanduc-JChemPhys-2017, Kanduc-PhysRevE-2018}. The dual-geometry consideration allowed the authors to find the ``line-tension-stiffness'' $\ppartial\kappa/\ppartial\theta$ correction, estimated as $\bigO(1/r)$, from the Tolman length and the apparent line tension of cylindrical droplets~\cite{Kanduc-JChemPhys-2017}. This correction was subsequently adopted with axisymmetric droplets~\cite{Kanduc-PhysRevE-2018}, implicitly assuming $\ppartial\kappaa/\ppartial\thetaa = \ppartial\kappac/\ppartial\thetac$. In the present work, we do not assume $\ppartial\kappaa/\ppartial\ra = \ppartial\kappac/\ppartial\rc$; however, these derivatives are shown \eqref{pdkappa_r-estimation} to be $\bigO \bigl( 1/r^2 \bigr)$, \mbox{i.\,e.}, negligible.

Finally, let us discuss if the proposed technique can be easily adapted to determine the \emph{size-dependent} line tension $\kappaa$ and/or $\kappac$ (not only their limit $\kappabndl$ at $r \to \infty$). Equation~\eqref{cYea-gYec} does not contain solid--fluid surface tensions, but it contains the surface tensions $\sgmabac$. They can be calculated separately for free spherical/cylindrical droplets of the same radius $\Rac$ of the liquid--gas surface as for the sessile droplet of the same type of geometry (thus, at the same value of chemical potential). However, since $\kappaa(\mu,\ra)$ and $\kappac(\mu,\rc)$ are different functions, the expression $\ppartial\kappaa/\ppartial\ra - \ppartial\kappac/\ppartial\rc$ on the right-hand side of Eq.~\eqref{cYea-gYec} is still $\bigO \bigl(\!(\delta\mu)^{2\,}\!\bigr) = \bigO \bigl( 1/\ra^2 \bigr)$ and therefore cannot be neglected even in the second order in $1/\ra$. Thus, employing equation~\eqref{cYea-gYec} is not a viable method for obtaining more detailed information on the line tension from dual-geometry measurements of $\cos\theta(1/r)$.

\begin{acknowledgments}
The authors thank \mbox{N.\,A.}~Volkov and \mbox{M.\,S.}~Polovinkin for valuable discussions and comments on the manuscript. The research has been carried out with financial support of the Russian Science Foundation, grant No.~22-13-00151, \href{https://rscf.ru/en/project/22-13-00151/}{https://rscf.ru/en/project/22-13-00151/}
\end{acknowledgments}

\appendix

\section{\label{appx:lt_drop_size_variable} Droplet size variable(s) the line tension depends on}

In Sec.~\ref{sec:equilibrium-cond}, the equilibrium conditions \eqref{Lea} and \eqref{gYea} (for axisymmetric droplets), and \eqref{Lec} and \eqref{gYec} (for cylindrical droplets) were derived, supposing the line tension $\kappaac$ to depend on the droplet size via its radius/half-width $\rac$ on the substrate. However, in the way these conditions were derived, this choice of the size variable(s) for the line tension is neither unique nor obvious.

What if we suppose the line tension to depend on the radius $R$ of the liquid--gas interface and/or the contact angle $\theta$? If the liquid--gas interface is still assumed to be spherical/cylindrical, derivation considered in Sec.~\ref{sec:equilibrium-cond} should be modified. For $\kappaac = \kappaac(\Rac,\thetaac)$, the equilibrium conditions \eqref{GTP-R-theta-eq_conds} applied to the functions $\Omegaac(\Rac,\thetaac)$ yield, for axisymmetric droplets,
\begin{gather*}
p^\alpha \! - p^\beta \! = \frac{2\sgmaba \!}{\Ra} + \Biggl[ \! \frac{\ppartial \kappaa}{\ppartial \thetaa} \frac{\sin\thetaa}{\Ra} - \frac{\ppartial \kappaa}{\ppartial \Ra} \cos\thetaa \! \Biggr] \frac{2 \sin\thetaa}{(1 - \cos\thetaa)^2 \Ra}, \label{gLea-Ra-thetaa} \\[0.5em]
\sgmaba \! \cos\thetaa = - \Dsgmg \! - \frac{\kappaa}{\ra} +  \frac{\ppartial \kappaa}{\ppartial \thetaa} \frac{2 + \cos\thetaa}{\Ra} - \frac{\ppartial \kappaa}{\ppartial \Ra} \frac{(1 + \cos\theta)^2 \!}{\sin\thetaa},  \label{gYea-Ra-thetaa}
\end{gather*}
and for cylindrical droplets,
\begin{gather*}
p^\alpha \! - p^\beta \! = \frac{\sgmabc \!}{\Rc} + \frac{ \dfrac{\ppartial \kappac}{\ppartial \thetac} \dfrac{\sin\thetac}{\Rc} - \dfrac{\ppartial \kappac}{\ppartial \Rc} \cos\thetac}{(\sin\thetac - \thetac \cos\thetac) \Rc}, \label{gLec-Rc-thetac} \\[1em]
\sgmabc \! \cos\thetac = - \Dsgmg + \frac{ \dfrac{\ppartial \kappac}{\ppartial \thetac} \dfrac{\thetac - \sin\thetac \cos\thetac}{\Rc} - \dfrac{\ppartial \kappac}{\ppartial \Rc} \sin^2\!\thetac}{\sin\thetac - \thetac \cos\thetac}.  \label{gYec-Rc-thetac}
\end{gather*}

As can be seen, the equilibrium conditions for $\theta$ (the generalized Young equations) and for $R$ (the Laplace equations) are subject to change and mix. The latter will only regain their original form at $ (\ppartial \kappa / \ppartial R) \cos\theta = R^{-1} (\ppartial \kappa / \ppartial \theta) \sin\theta$, a condition that corresponds to the case $\kappa = \kappa(r)$. In all other cases, the effective force that tends to change $R$ and/or $\theta$ due to dependence $\kappa(R,\theta)$ modifies the Laplace equation in both geometries. This nonlocal effect of the line tension has no reasonable explanation. The liquid--gas interface is not something rigid, and a force, tending to change it locally at the contact line, cannot affect the whole interface. The above equations are thus consequences of the assumption of a spherical/cylindrical shape for the interface.

Let us consider a variational derivation of the equilibrium conditions~\cite{Rusanov-Deform5-ColloidJUSSR-1977, Rusanov-SurfSciRep-1996, Tatyanenko-IPHT-2017, Tatyanenko-ColloidJ-2019}, wherein the droplet is not initially assumed to be a spherical/cylindrical segment, and then discuss, within this more general context, why $r$ is the only droplet-size variable on which the line tension can be assumed to depend.

\begin{figure*}[t]
\includegraphics[width=5.7cm]{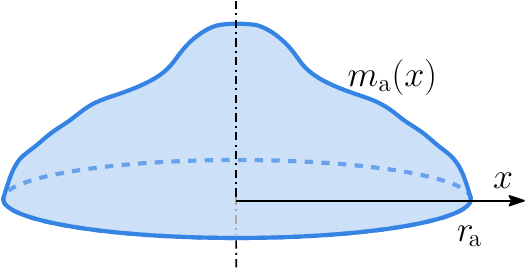}\hfill
\includegraphics[width=5.7cm]{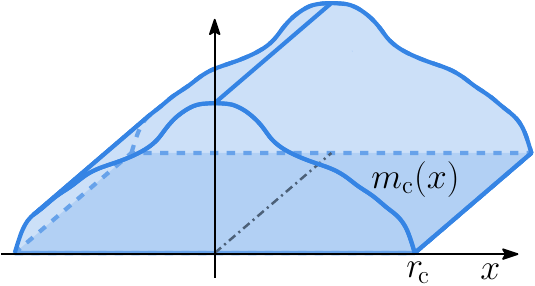}\hfill
\includegraphics[width=5.0cm]{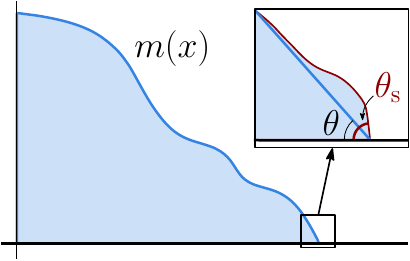}

\vspace{0.1cm}
\hfill (a) \hspace{6.2 cm} (b)  \hspace{5.5 cm} (c) \hfill\hfill
\caption{\label{fig:profiles-variational}Axisymmetric (a) and cylindrical (b) shapes of sessile droplets in variational consideration. (c) A local alteration in the [axisymmetric or cylindrical] profile of a droplet near its three-phase contact line, which can significantly change the contact-angle-dependent line tension, making an arbitrarily small contribution to the integral part of the functional~\eqref{genOmega-variation} of the variation $\delta\Omega$. The altered profile with the slope $m'(r) = - \tan\thetas$ (where $\thetas$: $\ppartial\kappa/\ppartial \theta |_{\theta = \thetas} = 0$) corresponds to an extremum of the nonintegral line term, which makes the $\delta m'\!(r)$ term equal to zero, while the $\delta r$ term remains zero since the position of the contact line does not alter.}
\end{figure*}

We assume that the axisymmetric droplet has an arbitrary radial profile, $\ma(x)$, where $x$ is the radial coordinate in cylindrical coordinates. Similarly, we assume that the cylindrical droplet (in this case, a segment of a curvilinear cylinder) has an arbitrary transversal profile, $\mc(x)$, where $x$ is the transversal Cartesian coordinate (\mbox{i.\,e.}, the coordinate along the substrate surface and perpendicular to the droplet's longitudinal axis). Refer to Figs.~\ref{fig:profiles-variational}(a) and \ref{fig:profiles-variational}(b) for illustration. The profiles $\mac(x)$ are assumed to be twice differentiable. The grand potential $\Omegaac$ of the system in both geometries can be expressed as a functional of $\mac(x)$:
\begin{align}
&\begin{aligned}
\Omegaac &[\mac(x)] = \int_0^{\rac} \! \omegaac \bigl( x, \mac(x), \mac'(x) \bigr) \dif x   \\
&{} + \Lac \kappaac \bigl(\rac,\mac'\!(\rac) \bigr) + \Omega^\gamma - p^\beta \Vtot + \sgmbg \Atot
\end{aligned} \label{genOmega-functional} \\[0.6em]
\intertext{with}
&\omegaac \equiv
\begin{dcases}
2\pi x \Bigl[ \Dsgmg\! + \sgmaba \! \sqrt{1 +  \smash[b]{(\ma'\!(x)\!)^2}} - \bigl( p^\alpha \! - p^\beta \bigr) \ma(x) \Bigr], \\[0.2em]
\Lc \Bigl[\Dsgmg\! + \sgmabc \! \sqrt{1 +  \smash[b]{(\mc'\!(x)\!)^2}} - \bigl( p^\alpha \! - p^\beta \bigr) \mc(x) \Bigr],
\end{dcases} \nonumber
\end{align}
where the symbol `$\prime$' indicates a derivative with respect to the argument of the function, $\Vtot = \Vaac + \Vbac = \const$, $\Atot = \Aagac + \Abgac = \const$. Here we take into account that
\begin{gather*}
\Vaa = 2 \pi \! \int_0^{\ra} \! x \, \ma(x) \dif x,
\quad
\Aaba = 2 \pi \! \int_0^{\ra} \! x \, \sqrt{1 + \smash[b]{(\mc'(x)\!)^2}} \dif x,\\
\Vac = \Lc \! \int_0^{\rc} \!\! \mc(x) \dif x,
\quad
\Aabc = \Lc \! \int_0^{\rc} \!\! \sqrt{1 + \smash[b]{(\mc'(x)\!)^2}} \dif x,\\
\Aaga = \pi \ra^2 =  2 \pi \! \int_0^{\ra} \!\! x \dif x,
\quad \!\!
\Aagc = \Lc \rc = \Lc \! \int_0^{\rc} \!\! \dif x,
\quad \!\!
\La = 2 \pi \ra,
\end{gather*}
and $\Lc$ does not depend on $\rc$.

In the functional~\eqref{genOmega-functional}, not only the droplet profile $\mac(x)$ is variable, but also the position $\rac$ of the right endpoint of the interval, as well as the nonintegral term $\Lac \kappaac$ with a variable $\La = 2\pi \ra$ for an axisymmetric droplet and a nonvariable $\Lc$ for a cylindrical droplet.  In the context of our analysis, we suppose the line tension $\kappaac$ to depend both on the radius/half-width $\rac$ of the droplet on the substrate and the contact angle $\thetaac$, \mbox{i.\,e.}, the local slope $\mac'$ of the profile at $x = \rac$, since $\mac'\!(\rac) = - \tan\thetaac$.

To find the equilibrium equations, we need to write down the first variation of the functional~\eqref{genOmega-functional} containing both integral and nonintegral contributions, with the right endpoint that can freely move along the $x$ axis. The boundary conditions for $\mac(x)$ are $\mac'\!(0)=0$ and $\mac(\rac)=0$.
Taking into account these boundary conditions, the first variation of the functional~\eqref{genOmega-functional} can be expressed as~\cite{Gelfand-CalcVariations-1963}
\begin{align}
&\delta\Omegaac \simeq \delta (\Lac\kappaac) + \! \int_0^{\rac} \!\! \left(\! \frac{\ppartial \omegaac}{\ppartial \mac} \delta m + \frac{\ppartial \omegaac}{\ppartial \mac'} \delta m' \! \right) \dif x  \nonumber \\
& {} + \omegaac \bigl( \rac, \mac(\rac), \mac'\!(\rac) \bigr) \, \delta \rac = \delta ( \Lac \kappaac)   \label{genOmega-variation} \\[0.2em]
& {} + \! \int_0^{\rac} \!\! \left( \! \frac{\ppartial \omegaac}{\ppartial \mac} - \frac{\dif}{\dif x} \frac{\ppartial \omegaac}{\ppartial \mac'} \! \right) \! \delta m \, \dif x + \left( \! \omegaac - \mac' \frac{\ppartial \omegaac}{\ppartial \mac'} \! \right) \! \Biggr|_{x=\rac} \!\!\!\!\!\!\!\!\!\!\!\!\!\!\! \delta \rac   \nonumber \\
\intertext{with $\delta (\Lac\kappaac)$ for the axisymmetric/cylindrical profile as}
&\delta (\Lac\kappaac) =
\begin{dcases}
2\pi\ra \! \left[ \left( \! \frac{\kappaa}{\ra} + \frac{\ppartial \kappaa}{\ppartial\ra} \! \right) \! \delta\ra - \frac{\ppartial \kappaa}{\ppartial\thetaa} \cos^2\!\thetaa \, \delta m'\!(\ra) \right]\!, \\[0.2em]
\Lc \! \left[ \frac{\ppartial \kappac}{\ppartial\rc} \, \delta\rc - \frac{\ppartial \kappac}{\ppartial\thetac} \cos^2\!\thetac \, \delta m'\!(\rc) \right]\!.
\end{dcases} \nonumber
\end{align}
Here $\ppartial\kappaac/\ppartial \mac'\!(\rac) = (\ppartial\kappaac/\ppartial \thetaac)(\dif \mac'\!(\rac)\!/\!\dif\thetaac)^{-1} \! = - (\ppartial\kappaac/\ppartial \thetaac) \cos^2\!\thetaac$  is used.

The variation of the droplet profile, $\delta m(x)$, can also be considered at a fixed position of the endpoint $r$, \mbox{i.\,e.}, at $\delta r = 0$. It is less clear that the variation of the profile slope at the endpoint, $\delta m'\!(r)$, should also be considered as independent. A slight alteration in the profile $m(x)$ within the vicinity of the endpoint can significantly modify $\delta m'\!(r)$, while maintaining a negligible change in the integral term in~\eqref{genOmega-variation} and keeping $\delta r = 0$ by fixing the position of the contact line, as depicted in Fig.~\ref{fig:profiles-variational}(c). Thus, the stationary conditions for the functional~\eqref{genOmega-functional}, $\delta\Omega = 0$, decompose into three equations~\cite{Rusanov-Deform5-ColloidJUSSR-1977, Rusanov-SurfSciRep-1996}:

I) The Euler--Lagrange equation $\ppartial\omega/\ppartial m = \dif (\ppartial\omega/\ppartial m')/\dif x$:
\begin{gather}
\left( \frac{\dif}
{\dif x} + \frac{1}{x} \right) \frac{\ma'}{\sqrt{1 + \smash[b]{(\ma')^2}}} = - \frac{p^\alpha - p^\beta}{\sgmaba}, \label{ELa} \\[0.3em]
\frac{\dif}{\dif x} \frac{\mc'}{\sqrt{1 + \smash[b]{(\mc')^2}}} = - \frac{p^\alpha - p^\beta}{\sgmabc}. \label{ELc}
\end{gather}
The left-hand side of each equation represents the sum of the principal curvatures of the droplet's free surface, which is constant and equal to $- \bigl( p^\alpha - p^\beta \bigr)/\sgmabac$. Since $\mc(x)$ is the \emph{transversal} profile and the principal curvature along the longitudinal coordinate is zero, the sum of the principal curvatures in Eq.~\eqref{ELc} equals $-1/\Rc$. Therefore, this equation is the Laplace equation~\eqref{Lec} for a circular-cylindrical droplet. For an axisymmetric droplet, the condition of the constant sum of the principal curvatures, in conjunction with the boundary condition $\ma'(0) = 0$, indicates that the surface is spherical. Consequently, the sum of its principal curvatures is $- 2/\Ra$, and Eq.~\eqref{ELa} is the Laplace equation~\eqref{Lea} for the axisymmetric (spherical) droplet.

II) The transversality condition (taking into account the freedom of the endpoint to move along the $x$ axis according to the boundary condition $m(r) = 0$) written as a condition of the zero coefficient at $\delta r$ in the expression~\eqref{genOmega-variation}:
\begin{gather*}
\Dsgmg + \sgmaba \! \sqrt{1 + \smash[b]{(\ma'(\ra)\!)^2}} + \frac1{\ra}
 \frac{\ppartial (\kappaa \ra)}{\ppartial \ra} = \frac{\sgmaba (\ma'(\ra)\!)^2}{\sqrt{1 + \smash[b]{(\ma'(\ra)\!)^2}}}, \\[0.5em]
\Dsgmg + \sgmabc \sqrt{1 + \smash[b]{(\mc'(\rc)\!)^2}} + \frac{\ppartial \kappac}{\ppartial \rc} = \frac{\sgmabc (\mc'(\rc)\!)^2}{\sqrt{1+ \smash[b]{(\mc'(\rc)\!)^2}}}.
\end{gather*}
Since we have already derived the Laplace equations~\eqref{Lea} and~\eqref{Lec} meaning that the droplets' free surfaces are spherical and cylindrical segments, respectively, $\mac'(\rac) = - \tan \thetaac$; thus, it can be demonstrated that the equations above yield the generalized Young equations \eqref{gYea} and \eqref{gYec}.

III) The condition of the zero coefficient at $\delta m'\!(r)$ yields $\ppartial\kappa / \ppartial m'\!(r) = 0$, \mbox{i.\,e.}, $\ppartial\kappa / \ppartial \theta = 0$. Given that the preceding two equations determine a spherical/cylindrical segment and the values of its two parameters, this additional equation makes the entire set of equations generally inconsistent.

Let us discuss again the question of the independence of the variation $\delta m'\!(r)$ from the variation of the integral in~\eqref{genOmega-variation}. This presumed independence led to conditions (I) and (III) in their particular form and made them independent (in contrast to the approach used before, including the beginning of this section). However, this assumption apparently contradicts the ``regularity'' of the function $m'\!(x)$ at $x = r$. The dependence of the line tension $\kappa$ on the contact angle $\theta$ leads to the common functional $\Omega[m(x)]$ being dependent on the derivative of the function $m(x)$ \emph{at a single point} $x = r$ (\mbox{i.\,e.}, the functional has a \emph{singular} part similar to the derivative of the Dirac delta function). This generally makes the variational problem ill posed, in the sense that it lacks a solution within the domain of continuously differentiable functions. It is quite simple to construct a sequence of profiles $\{m_i(x)\}_{i=1}^\infty$, only slightly altering the regular one (given by the solution of the Laplace and the generalized Young equations) in a very small vicinity of $x = r$ in such a way that $m'_i(r) \xrightarrow[i \to \infty]{} - \tan\thetas$, where $\thetas$: $\ppartial\kappa/\ppartial \theta |_{\theta = \thetas} = 0$ (see Fig.~\ref{fig:profiles-variational}(c)). However, such a sequence does not generally converge to a regular function, instead producing a singular contribution, clearly nonphysical in nature, to the equilibrium profile $m(x)$. A reasonable regularization procedure would involve the elimination of the singular part of the function, which corresponds to ignoring the condition (III) based on the contact-angle dependence of the line tension. Thus, the only case in which the problem is both physically meaningful and mathematically well posed is when $\ppartial\kappa/\ppartial \theta \equiv 0$.

In regard to the possible dependence of the line tension $\kappaac$ on the curvature radius $\Rac$ of the liquid--gas interface, its consideration should lead to similar conclusions. Assuming such a dependence on the \emph{local} value of the interface curvature/radius at the contact line, a similar construction with a local alteration of the droplet profile can be made, leading to the same conclusion. However, if the line tension is assumed to depend on the [nonlocal] mean curvature of the liquid--gas interface, the rigorous variational problem statement remains unclear. This is because it is generally not known in advance that the solution will be a profile $\mac(x)$ of constant curvature. If this curvature is \emph{assumed} to be constant and the line tension depends on \emph{its} value, this would only affect the variation $\delta(\Lac\kappaac)$ in~\eqref{genOmega-variation}, and specifically, the value of the line tension, but not the generalized Young equation for the contact angle. The curvature is still determined by the value of the chemical potential via the Gibbs--Duhem and the Laplace equations (which will be a consequence of the Euler--Lagrange equation). Therefore, the variation $\delta(\Lac\kappaac)$ is to be considered at a constant liquid--gas interface curvature, which yields the same outcome as if the curvature dependence of the line tension were incorporated directly into its chemical potential dependence.

\bibliography{LT-DG}

\end{document}